\documentclass[sigconf]{acmart}

\copyrightyear{2019}
\acmYear{2019}
\setcopyright{acmlicensed}
\acmConference[Dublin '19]%
  {Dublin '19: 6th ACM International Conference on Nanoscale Computing and Communication}
  {September 25--27, 2019}
  {Dublin, Ireland}

\newcommand{\Ec}{\mathcal{E}}

\begin{document}

\title{Graphene epsilon-near-zero plasmonic crystals}

\author{Marios Mattheakis}
\email{mariosmat@g.harvard.edu}
\orcid{0000-0001-6450-1128}
\affiliation{%
  \department{School of Engineering and Applied Sciences}
  \institution{Harvard University}
  \city{Cambridge}
  \state{MA, USA}
}

\author{Matthias Maier}
\email{maier@math.tamu.edu}
\orcid{0000-0002-4960-5217}
\affiliation{%
  \department{Department of Mathematics}
  \institution{Texas A\&M University}
  \city{College Station}
  \state{TX, USA}
}

\author{Wei Xi Boo}
\affiliation{%
  \institution{University of Minnesota}
  \city{Minneapolis}
  \state{MN, USA}
}

\author{Efthimios Kaxiras}
\affiliation{%
  \department{School of Engineering and Applied Sciences}
  \department{Department of Physics}
  \institution{Harvard University}
  \city{Cambridge}
  \state{MA, USA}
}

\renewcommand{\shortauthors}{Mattheakis et al.}

\begin{abstract}

  Plasmonic crystals are a class of optical metamaterials that consist of
  engineered structures at the sub-wavelength scale. They exhibit optical
  properties that are not found under normal circumstances in nature, such as
  negative-refractive-index and epsilon-near-zero (ENZ) behavior. 
  Graphene-based plasmonic crystals present
  linear, elliptical,   or hyperbolic dispersion relations that exhibit ENZ behavior, normal or
  negative-index diffraction.   The optical properties  can be dynamically tuned by controlling the operating frequency and the doping  level of graphene. 
We propose a construction approach to expand the frequency range of the ENZ behavior. We   demonstrate how the combination of a host material with an optical
  Lorentzian response in combination with a graphene conductivity that
  follows a Drude model leads to an  ENZ condition spanning a
  large frequency range.
  
\end{abstract}

\keywords{Graphene, metamaterials, plasmons}

\maketitle

\section{Introduction}

The {control of} electromagnetic (EM) properties {of optical materials on the nanoscale} has opened the door {for designing} novel devices and applications that include  nano-antennas with extremely short wavelength resonance \cite{nanoLet10_2010}, optical holography \cite{nanoLet15_2015}, and wireless nanocommunication \cite{ieee2015_grapheneAntennas,HASSAN20191}.
Controlling the structure of materials at the sub-wavelength scale
{enables the design of} metamaterials {that possess} properties
that are not found under normal circumstances in nature such as negative
refraction \cite{nature522_2015}  and epsilon-near-zero (ENZ) behavior \cite{prl97_2006}. One of the key ingredients
for designing such metamaterials is the creation of EM waves with much
shorter wavelengths than those of the incident light. These sub-wavelength
waves involve electronic motion (plasmons) coupled with EM surface waves
and are referred to as surface plasmon polaritons \cite{prb80_2009,natPhot6_2012}.

Plasmonic crystals are {a class of metamaterials that is of}
particular interest. They consist of metallic layers arranged
periodically with sub-wavelength distance in a dielectric host. This class
of metamaterials offers {fine} control of EM properties and can serve
as ENZ \cite{mattheakis2016,maierPRB2018} and negative refractive index media \cite{wang2012}.
ENZ metamaterials exhibit properties which cannot be obtained by
traditional photonic systems. These features include wave propagation with
no phase delay and diffraction, decoupling of spatial and temporal field
variations, propagation through very narrow channels, and ultra fast phase
transitions  \cite{prl97_2006, mattheakis2016}. Many novel {optical} devices have been proposed indicating the broad prospects of the ENZ effect \cite{natMat10_2011,natPhot9_2015,natPhot12_2018_ENZnanoantenna}.
{However, novel metals have the unfortunate side effect that
their use in plasmonic platforms} restricts {the frequency range where}
above exotic optical effects occur. In addition, {the resulting
plasmonic device regularly} suffers from high optical losses {and}
construction defects yield non perfect planar layers which also negatively affect the
optical properties. {All of these
issues} may no longer be critical due to the discovery of truly
two-dimensional (2D) materials, such as graphene, black phosphorus,
hexagonal boron nitride, and molybdenum disulfide.
They promise a new era of control of optical properties of photonic and plasmonic
devices \cite{PhysRevLett.113.106802,natNano10_2015,prl116_2016,natPhot8_2014,Valagiannopoulos_2017}. {In particular, graphene} supports plasmons with
ultra-sub-wavelength behavior, ranging from terahertz (THz) to infrared
frequencies. Due to the confining of electrons in two dimensions, the
quantum effects are very pronounced and thus 2D materials pose exceptional
optical properties and high quantum efficiency for light-matter
interaction  \cite{prb80_2009,natPhot6_2012}.

Graphene is a quite interesting material because {it allows} control
{of} electronic and, in turn, plasmonic properties by changing
the density of the free charge carriers through an external gate field,
chemical deposition, or intercalation \cite{natPhot6_2012,natPhot8_2014,PhysRevB.97.195435}. It has been
shown that graphene plasmonic crystals {can serve as an ingredient in}
a tunable metamaterial \cite{mattheakis2016,maierPRB2018,wang2012}. {In particular,  a plasmonic crystal consisting of} periodically stacked
graphene layers {embedded} in a dielectric host leads to negative
refraction and ENZ behavior in the THZ and infrared frequencies, {and
allows to tune} optical properties by controlling the operation frequency
or the doping amount \cite{mattheakis2016}. It has been furthermore shown that
the ENZ condition coincides with a \emph{plasmonic Dirac cone} in the
wavenumber space, where the plasmonic Dirac cone is formed by two
{asymptotically} linear, intersecting dispersion bands \cite{mattheakis2016,maierPRB2018,natMat10_2011,natPhot9_2015}.
{Such a graphene-based plasmonic crystal design, however, has the major
shortcoming that ENZ behavior can only be observed for a very small
frequency range. This is due to the fact that a plasmonic crystal with a
dielectric host that has frequency independent optical properties  admits
only a single plasmonic Dirac point which restricts the ENZ effect to a
very narrow frequency range. This is  disadvantageous for a number of
optical applications involving a continuous spectrum of EM waves.
}
In this letter we introduce a further degree of freedom in controlling the
optical properties of graphene plasmonic crystals by using  Lorentz
dispersive dielectric host materials \cite{solidStateBook}. The dielectric function of this family
of materials depends on the operating frequency. We find that by
using an appropriate Lorentz host we obtain multiple plasmonic Dirac points
resulting in a wide frequency range exhibiting an ENZ effect. {For a
particular host material that we examine, Magnesium Oxide, we demonstrate
that (a) two Dirac points emerge in the THz band; and (b) that for a large
frequency interval containing the two Dirac points the ENZ condition is
fulfilled.}
{As a second design example,} we propose a hypothetical dielectric with
two Lorentz resonances leading to three Dirac points and {an even wider} frequency spectrum of 
ENZ behavior. {This example is motivated by the significant progress
that was made in (atomistic) material engineering recently,
that has the prospect of enabling the engineering of materials with
desirable electronic properties.} For instance, Van der Waals
heterostructures are comprised of different 2D materials that are stacked
in a particular way  to provide desirable electronic and optical features \cite{nature499_2013,Novoselovaac9439}.

The remainder of this paper is organized as follows. In Section 2 we review
the graphene-based plasmonic crystal by using Bloch theory and discuss the
mathematical relation between the plasmonic Dirac point and ENZ effect. In
Section 3 we extend the frequency range of the ENZ effect by introducing
Lorentz materials as the ambient material of the graphene layers. The paper
concludes in Section 4 with a summary of the key ideas introduced in this
{letter and an outlook.}

\section{Bloch-wave Theory}

We consider a plasmonic crystal that consists of a
dielectric {host surrounding} flat in which periodic 2D metallic sheets
that are parallel to the $yz$-plane located at $x=n d$, for integer
$n$. {Here, $d$} denotes the structural period. The metallic sheets are
described by an isotropic surface conductivity $\sigma$. The dielectric
host is considered to be an anisotropic material described by a uniaxial
dielectric tensor $\text{diag}(\varepsilon_x,\varepsilon_y, \varepsilon_z)$
with out-of-plane component $\varepsilon_x=\text{const}$ and $x$-dependent
in-plane components $\varepsilon_y(x)=\varepsilon_z(x)$; the vacuum
permittivity is set to unity, $\varepsilon_0=1$. We examine
transverse-magnetic (TM) time-harmonic waves with electric and magnetic
field components ${\boldsymbol E}=(E_x, 0, E_z)$ and ${\boldsymbol H} = (0,H_y,0)$,
respectively. Imposing the rotational symmetry of the crystal and the TM
polarization in Maxwell equations yield fields which are translation
invariant along $y$ direction with guided modes align along $z$ direction \cite{mattheakis2016},
hence
\begin{align*}
  E_z(x,z) = \Ec(x) e^{ik_z z}\;,
\end{align*}
where $k_z$ is the wavenumber along  $z$ direction. The above assumptions
allow us to simplify the system of time dependent Maxwell equations  to a
one dimensional eigenvalue problem \cite{maierPRB2018}  with the Helmholtz type
governing equation
\begin{align}
  \label{eq:E_eq}
  \left[\frac{\partial^2 }{\partial x^2} - \kappa(k_z)
  \varepsilon_z(x)\right] \Ec(x) = 0,
  \quad
  \kappa(k_z) = \frac{k_z^2 - k_0^2\varepsilon_x}{\varepsilon_x},
\end{align}
where $k_0 = \omega/c$, $\omega=2 \pi f$ is the angular operating frequency and $f$ denotes the ordinary operating frequency, $c$ is the
free space speed of light. The metallic 2D sheets carry a surface current
$J_s = \sigma \Ec$ that acts as a boundary. Maxwell equations dictate that
on the boundaries  that are defined by the metallic sheets at $x=n d$, the
tangential electric field $\Ec$ must be continuous with discontinuous
derivative with a jump discontinuity due to the surface currents on the
metallic plates, hence the transmission conditions are \cite{maierPRB2018}:
\begin{align}
  \label{eq:BCcont}
  \begin{cases}
    \begin{aligned}
      \Ec^+ - \Ec^- &= 0,
      \\
      \frac{\partial }{\partial x}\Ec^+ - \frac{\partial }{\partial x}\Ec^- &=
      \frac{i \sigma}{\omega}\kappa(k_z)\Ec^+,
    \end{aligned}
  \end{cases}
\end{align}
where $(.)^\pm$ denotes the limit above $(+)$ and below $(-)$ of a metallic
boundary. We can get a closed system of equations by using the Bloch-wave
ansatz \cite{mattheakis2016,maierPRB2018} in the $x$ direction with $k_x$ indicating the real Bloch number:
\begin{align}
  \label{eq:bloch}
  \Ec(x) = e^{ik_x d} \Ec(x-d).
\end{align}
The combination of the transmission conditions (\ref{eq:BCcont}) and Bloch
ansatz (\ref{eq:bloch}) yields a closed system of boundary conditions,
viz.,
\begin{equation}
\label{eq:BCsys}
\begin{bmatrix}\Ec(d^-) \\ \Ec'(d^-)\end{bmatrix} =
  e^{i k_x d}
  \begin{bmatrix}
    1 & 0 \\
    - i \sigma/\omega\kappa(k_z)& 1
  \end{bmatrix}
  \begin{bmatrix}\Ec(0^+) \\ \Ec'(0^+)\end{bmatrix}.
\end{equation}
{Here, $\Ec'$ denotes the derivative of $\Ec$ with respect to $x$. In
order to solve the differential equation \eqref{eq:E_eq} with boundary
conditions \eqref{eq:BCsys} we introduce the two fundamental solutions
$\Ec_{(1)}$ and $\Ec_{(2)}$ of \eqref{eq:E_eq} that fulfill the initial
conditions:}
\begin{align*}
  \Ec_{(1)}(0) = 1, \; \Ec'_{(1)}(0) = 0, \quad
  \Ec_{(2)}(0) = 0, \; \Ec'_{(2)}(0) = 1.
\end{align*}
By substituting a linear combination $\Ec=c_1\,\Ec_{(1)}+c_2\,\Ec_{(2)}$
into (\ref{eq:BCsys}) we obtain
\begin{align*}
\Bigg( \begin{bmatrix}  \Ec_{(1)}(d) & \Ec_{(2)}(d) \\ \Ec'_{(1)}(d) &  \Ec'_{(2)}(d) \end{bmatrix}  -    e^{i k_x d}
 \begin{bmatrix}
      1 & 0 \\
      - i (\sigma/\omega)\kappa(k_z)& 1
    \end{bmatrix}        \Bigg) \begin{bmatrix}
    c_1 \\ c_2
\end{bmatrix}     = 0.
\end{align*}
This system admits a nontrivial solution (with $c_1\not=0$, or $c_2\not=0$)
whenever
\begin{multline}
  \label{eq:generalDispersion}
  D[{\boldsymbol k}]  = \text{det} \Bigg( \begin{bmatrix}  \Ec_{(1)}(d) & \Ec_{(2)}(d) \\ \Ec'_{(1)}(d) &  \Ec'_{(2)}(d) \end{bmatrix}
  \\
  -    e^{i k_x d}
  \begin{bmatrix}
    1 & 0 \\
    - i (\sigma/\omega)\kappa(k_z)& 1
  \end{bmatrix}
  \Bigg)= 0.
\end{multline}
{Equation $\eqref{eq:generalDispersion}$ now forms a dispersion
relation between $k_x$ and $k_z$.} It is convenient to work in a 2D
wavenumber space ${\boldsymbol k} = (k_x,k_z)$. In Ref. \cite{maierPRB2018}
formula (\ref{eq:generalDispersion}) is used to determine the dispersion
relation for several examples of dielectric functions, $\varepsilon_z(x)$,
including constant, parabolic, double-well, and nonsymmetric dielectric
profiles.
The plasmonic crystal has the character of an ENZ medium for values
$\boldsymbol k$ in the vicinity of the Dirac point ${\boldsymbol k}^* =
(k_x^*,\ k_z^*) = (0,\pm k_0\sqrt{\varepsilon_x})$ \cite{maierPRB2018}, that occurs
at the center of the Brillouin zone \cite{mattheakis2016}.
It has been shown that the occurence of the plasmonic Dirac point is a
universal behavior of 2D plasmonic crystals with any spatial dependence on
the dielectric host along the $x$ direction \cite{maierPRB2018}.

At ${\boldsymbol k}^*$ the effective in-plane permittivity component,
$\varepsilon_z^\text{eff}$, of the metamaterial becomes zero, hence in the
neighborhood of ${\boldsymbol k}^*$ we observe ENZ behavior.
{For the purpose of examining ENZ \emph{frequency
bands} in the following, we make the heuristic definition of speaking of an
ENZ behavior whenever $\big|\varepsilon_z^\text{eff}\big|<0.25$.}
{In this letter we focus on the frequency-dependent response of a
plasmonic crystal assuming that the host material is described by a
{ constant in space but possibly $\omega$ dependent permittivity}. Thus, as a
last preparatory step we solve the dispersion relation explicitly for the
special case of spatially constant $\varepsilon_x$ and $\varepsilon_z$. In
this case the fundamental solutions of the differential equation
\eqref{eq:E_eq} are given by}
\begin{align*}
  \Ec_{(1)} = \cosh\left(\sqrt{\kappa \varepsilon_z} x \right),
  \quad
  \Ec_{(2)} = \frac{1}{\sqrt{\kappa\varepsilon_z}}
  \sinh\left(\sqrt{\kappa \varepsilon_z} x \right).
\end{align*}
Substituting both solutions into dispersion relation
(\ref{eq:generalDispersion}) yields
\begin{align}
  \label{eq:ezConst_dispersion}
  \cos(k_x d) = \cosh(\sqrt{\kappa\varepsilon_z}d) - \frac{\xi_0
  \sqrt{\kappa\varepsilon_z}}{2}\sinh(\sqrt{\kappa\varepsilon_z} d),
\end{align}
where
\begin{align}
  \label{eq:xio}
  \xi_0 = -\frac{i \sigma}{\omega \varepsilon_{z}}
\end{align}
is the \emph{plasmonic thickness} that determines the {length scale of
plasmonic structures} perpendicular to the 2D material sheets \cite{mattheakis2016,maierPRB2018,wang2012}. For $\boldsymbol k$ close to $\boldsymbol
k^\ast$ the dispersion relation
\eqref{eq:ezConst_dispersion} reduces to
\begin{align*}
  \frac{k_x^2}{\varepsilon_z^\text{eff}} + \frac{k_z^2}{\varepsilon_x}
  = k_0^2.
\end{align*}
This implies that up to first order the effective permittivity
$\varepsilon_z^{\text{eff}}$ is given by
\begin{align}
  \label{eq:eps_eff}
  {\varepsilon_z^\text{eff}} = \left(1 -\frac{\xi_0}{d}\right){\varepsilon_z}.
\end{align}
{It can be shown by a refined argument that equation \eqref{eq:eps_eff}
is universal \cite{maier_2019}: An EM wave traveling through the plasmonic
crystal sees an \emph{effective} homogeneous medium with effective
permittivity $\varepsilon = \text{diag}\left( \varepsilon_x,
\varepsilon_z^\text{eff}, \varepsilon_z^\text{eff}\right)$. Therefore,
three distinct cases of wave propagation are possible: elliptic propagation
($\Re\,\varepsilon_z^\text{eff}>0$), hyperbolic propagation
($\Re\,\varepsilon_z^\text{eff}<0$), and ENZ behavior
($\Re\,\varepsilon_z^\text{eff} \approx 0$). The elliptical
dispersion bands yield propagating waves with normal diffraction, whereas
the hyperbolic bands correspond to negative refraction. The ENZ case is
described by a \emph{linear} Dirac dispersion and corresponds to
propagation without dispersion and phase delay \cite{mattheakis2016}.}

\begin{figure}
  \centering
  \includegraphics[scale=.23]{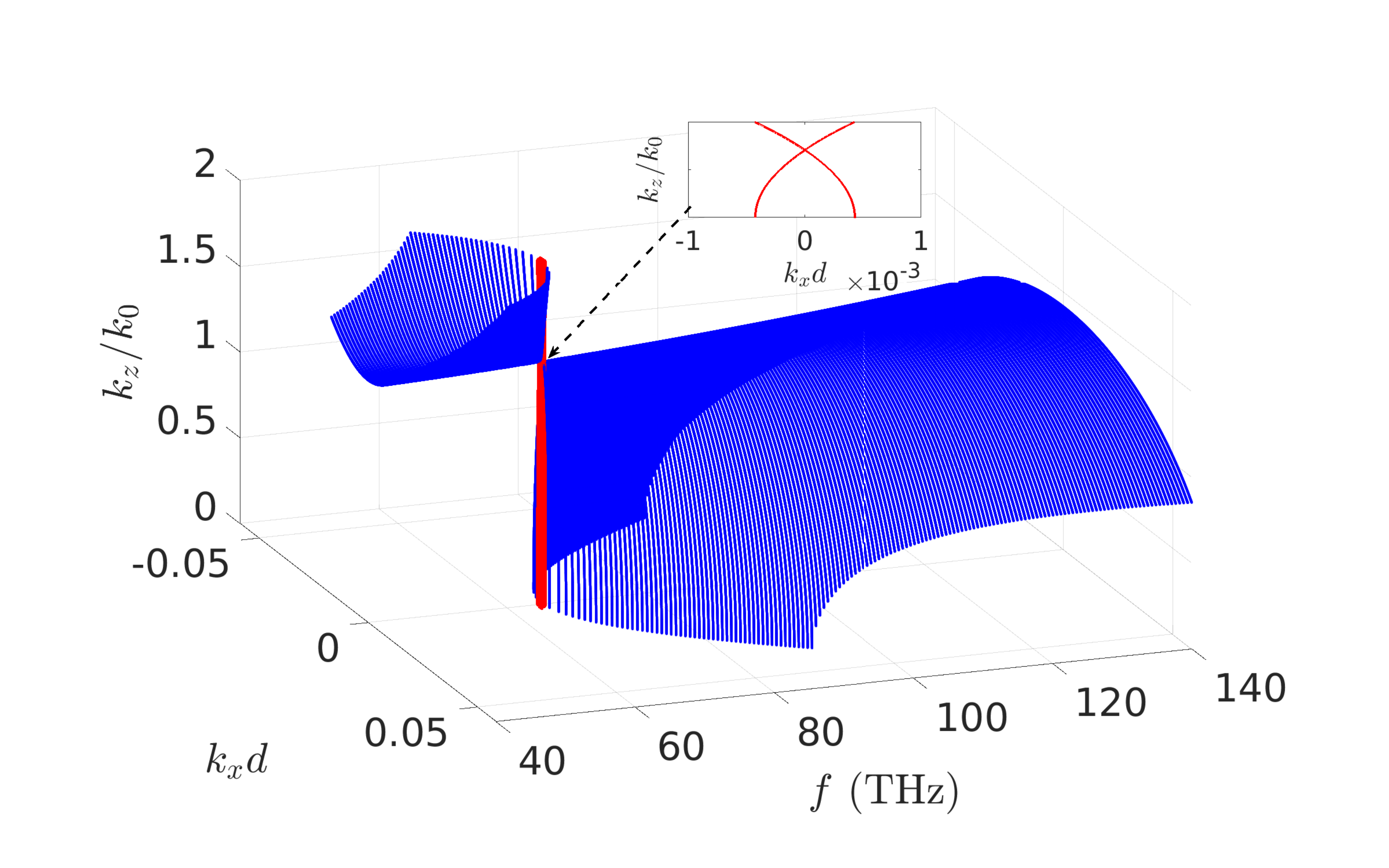}
  \caption{    Dispersion relation $k_x(k_z)$ as function of frequency for a graphene
    plasmonic crystal with silica glass as dielectric host. The Dirac cone is
    marked in red and also shown in inset.}
       \label{fig:bands_simple}
\end{figure}

Equation (\ref{eq:eps_eff}) provides a systematic way to design plasmonic
crystals with ENZ properties. We observe that we obtain
$\varepsilon_z^\text{eff}=0$ when $\xi_0=d$, hence for a given $d$ we can
tune the plasmonic thickness, $\xi_0$, to be equal to $d$, and vice versa.
{This also explains why graphene is such an exceptional plasmonic
platform. The surface conductivity of graphene depends on the
{operating} frequency as well as the doping,} hence we can dynamically
tune the surface conductivity $\sigma$ to achieve $\xi_0 \approx d$. More
precisely, the surface conductivity of graphene is given by Kubo's formula
which includes both interband and intraband electronic transitions.
Nevertheless, in THz and infrared frequencies the intraband transitions
dominate and thus, Kubo's formula is simplified to Drude model \cite{prb80_2009,natPhot6_2012}:
\begin{align}
  \label{eq:grapheneCond}
  \sigma(\omega) = \frac{i e^2 E_F}{\pi \hbar^2(\omega + i/\tau)},
\end{align}
where $E_F$ is the Fermi level associating with the density of electric
carriers and corresponding to the electronic doping; the $E_F$ can be
dynamically tuned by an external gate field or can be fixed by chemical
deposition  or intercalation \cite{natPhot6_2012,natPhot8_2014,PhysRevB.97.195435}.  The   electron charge is
denoted by $e$, $\hbar$ is the reduced Planck constant, and $\tau$ is
electronic relation time accounting for optical losses with {a
typically value of $\tau=0.5$ ps}.

{As a first example we} study a plasmonic crystal that consists of
doped graphene layers in a silica glass host with constant and equal
permittivity components of $\varepsilon_x = \varepsilon_z = 2$. In
Figure~\ref{fig:bands_simple} we show the dispersion relation $k_x(k_z)$
calculated by equation (\ref{eq:ezConst_dispersion}) for different values of
frequency $f$ in the range $[40,~ 140]$ THz. The structural period and
the Fermi level are kept constant at $d=20$ nm and $E_F=0.5$ eV,
respectively. We observe one plasmonic Dirac point that appears  at
$f=65$ THz (highlighted by red and shown in inset diagram of Figure~\ref{fig:bands_simple}), where the dispersion bands change from elliptical to hyperbolic.

\section{Lorentz host material}

{In the previous section we saw how a (single) plasmonic Dirac point
can appear in the dispersion of a 2D metal-based plasmonic crystal that
then---in the vicinity of the Dirac point---behaves as an ENZ medium. The
ENZ effect is, however, restricted to} a narrow range of frequencies. This
{is an issue for real applications that typically involve a range of
operating frequencies forming a wavepacket. We now introduce a Lorentz host
material into the plasmonic crystal. We demonstrate how this leads to
multiple Dirac points in the dispersion relation and consequently a
significantly expanded frequency range with ENZ behavior.}

{The Drude model takes only intraband electronic transition into
account that model free electrons in a metal response to an EM field.
Dielectric materials, however, do not have free charge carriers
(electrons). Here, the optical response due to an external EM field is
mainly caused by bound charges in the dielectric material where radiation
can be absorbed due to interband electronic transitions.} The Lorentz model
is a {phenomenological model} that takes {these interband
electronic transitions into account.} Dielectrics described by this model
are called \emph{Lorentz dispersive materials} \cite{solidStateBook}. The contribution of the
interband transition leads to a frequency dependent permittivity that reads
\begin{align}
  \label{eq:lorentzEps}
  \varepsilon(\omega) = \varepsilon_\infty+
  \frac{(\varepsilon_s-\varepsilon_\infty)\omega_0^2}{\omega_0^2-\omega^2
  +i\Gamma \omega }.
\end{align}
{Here,} $\omega_0$ denotes the resonance (natural) angular frequency of the
bound electrons, $\Gamma$ is the electron damping rate,
$\varepsilon_\infty$ and $\varepsilon_s$ are the high and low frequency
limits of the permittivity, respectively.

\begin{figure}[ht]
  \centering
  \includegraphics[scale=.23]{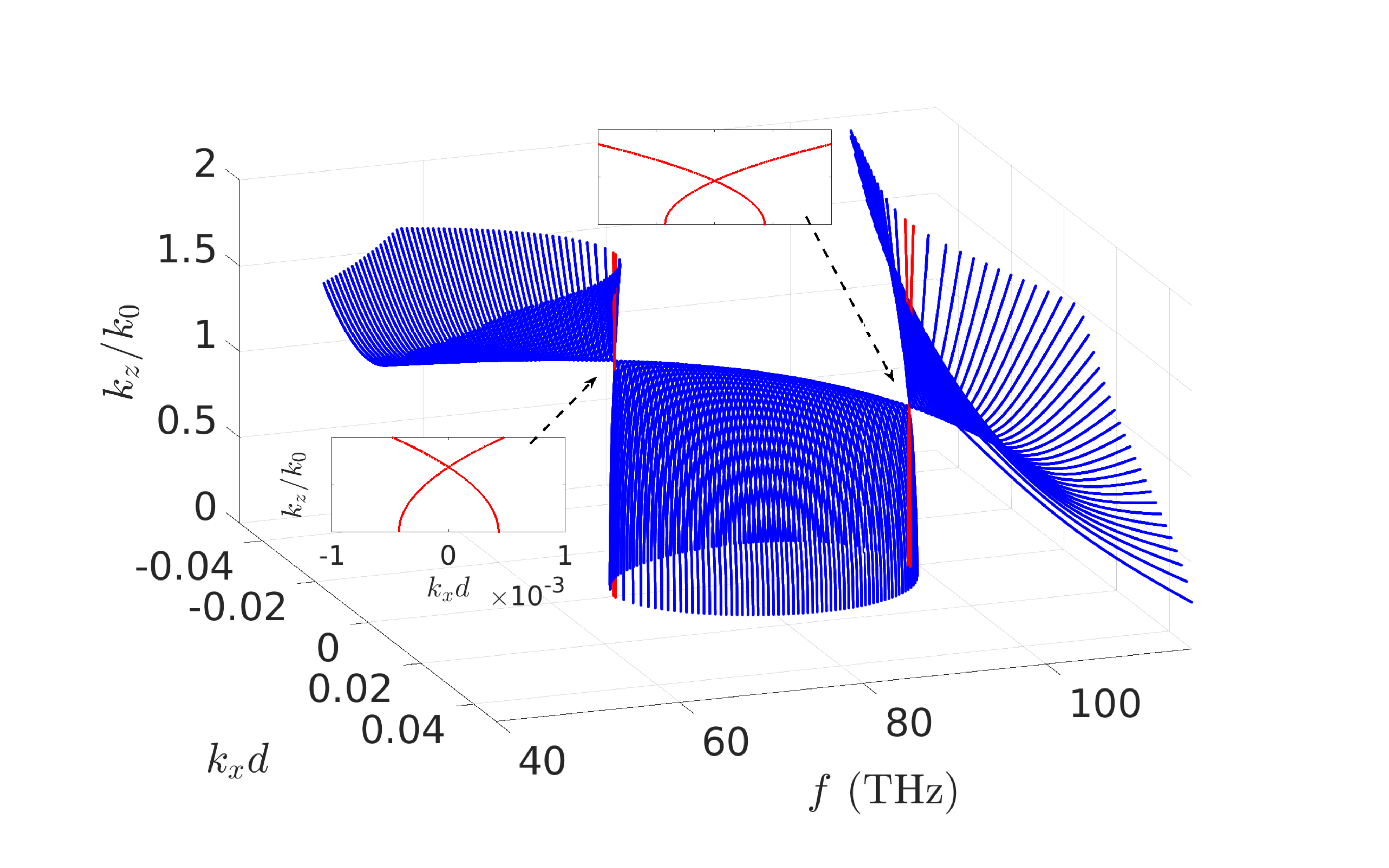}
    \includegraphics[scale=0.35]{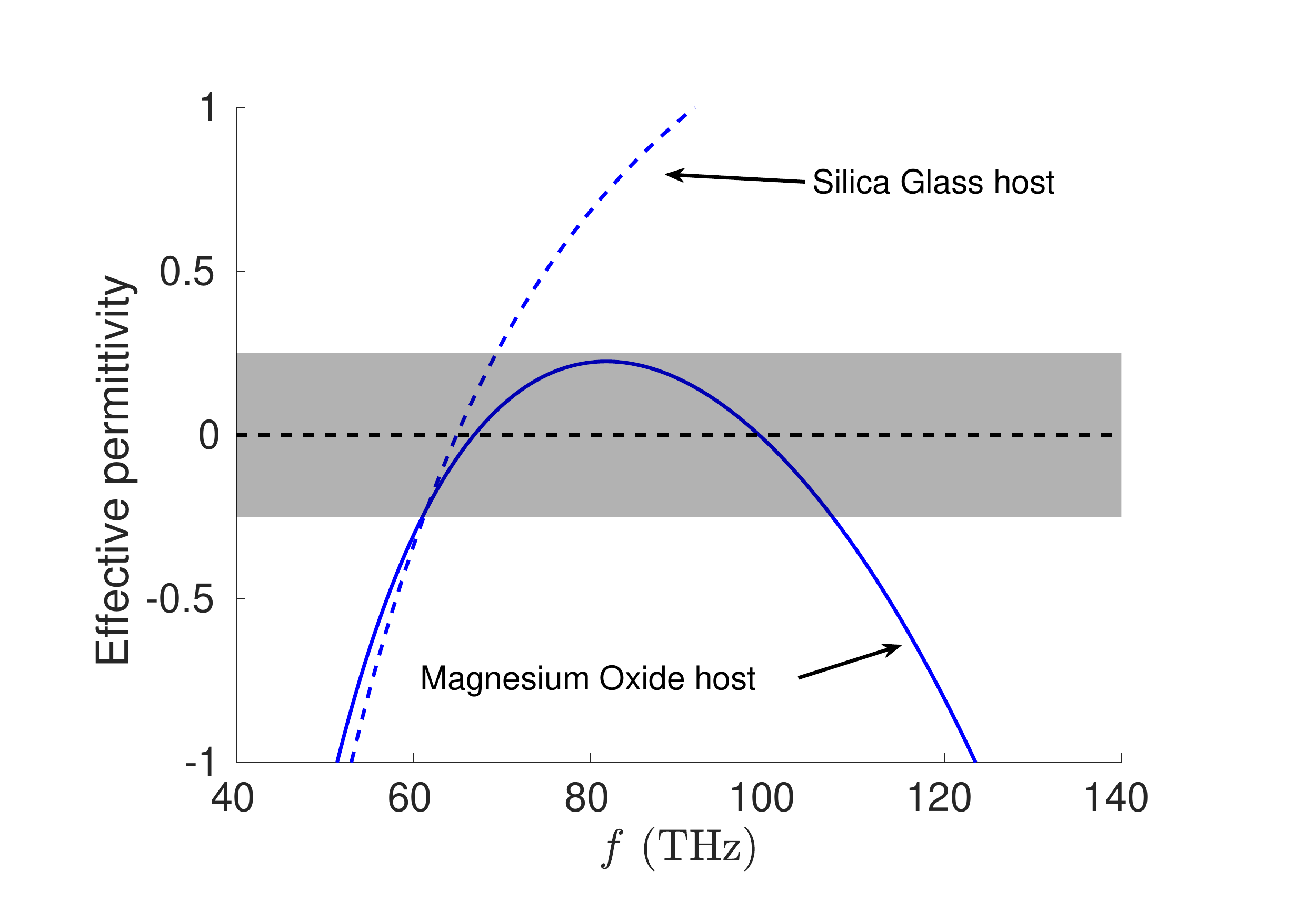}
  \caption{%
  	 	Upper:    Dispersion relation $k_x(k_z)$ as a function of frequency  for a graphene plasmonic crystal with Magnesium Oxide dielectric host; the two Dirac cones are marked in red and shown in the inset. Lower: Effective dielectric permittivity in frequency for a graphene plasmonic  crystal with Magnesium Oxide (solid) and silica glass (dashed) as host    materials; the shaded are represents the ENZ range.}
  \label{fig:bands_lorentz}
\end{figure}

{A concrete example for a plasmonic crystal with a Lorentz dispersive host
material that shows an extended ENZ band is Magnesium Oxide (MO).} MO's
parameters for the Lorentz model read: $\varepsilon_\infty = 11.2$,
$\varepsilon_s=2.6$, $\omega_0 = 1$ eV ($f=241.8$ THz), and $\Gamma=0$
\cite{mgo_notes}. We point out that MO is considered to be a lossless
dielectric and hence can be used for optical devices.
{We compute the dispersion relation (\ref{eq:ezConst_dispersion}) of the
plasmonic crystal for this case by setting the host permittivity to
$\varepsilon_z=\varepsilon_x=\varepsilon(\omega)$. The resulting dispersion relation
involving $(k_x,k_z)$ and the operating frequency $f$ are shown in 
Figure~\ref{fig:bands_lorentz} (upper). We indeed observe two Dirac points at
frequencies $f_1 = 67$ THz and $f_2 = 99$ THz. The two Dirac cones are marked in
red in the diagrams and pointed out in the inset of Figure \ref{fig:bands_lorentz}.} 

\noindent Since we are interested in the ENZ range, we calculate the
in-plane effective dielectric permittivity given by equation \eqref{eq:eps_eff}. The
results are shown in the lower panel of Figure~\ref{fig:bands_lorentz}. In particular, we compute the
$\varepsilon_z^\text{eff}$ as a function of the frequency {for our
example of a plasmonic crystal with MO host.}
For the sake of comparison we plot the
$\varepsilon_z^\text{eff}$ for a plasmonic crystal that was studied in
previous sections, that it, with  silica glass as dielectric host  instead of MO. The shaded area represents the ENZ range ($\left|\varepsilon_z^\text{eff}\right|<0.25$).
{We observe that indeed the ENZ band of the plasmonic crystal with Lorentz
dispersive host is significantly extended, about four times, compared to the previous example of a constant permittivity host material.}

\begin{figure}
  \centering
  \includegraphics[scale=.23]{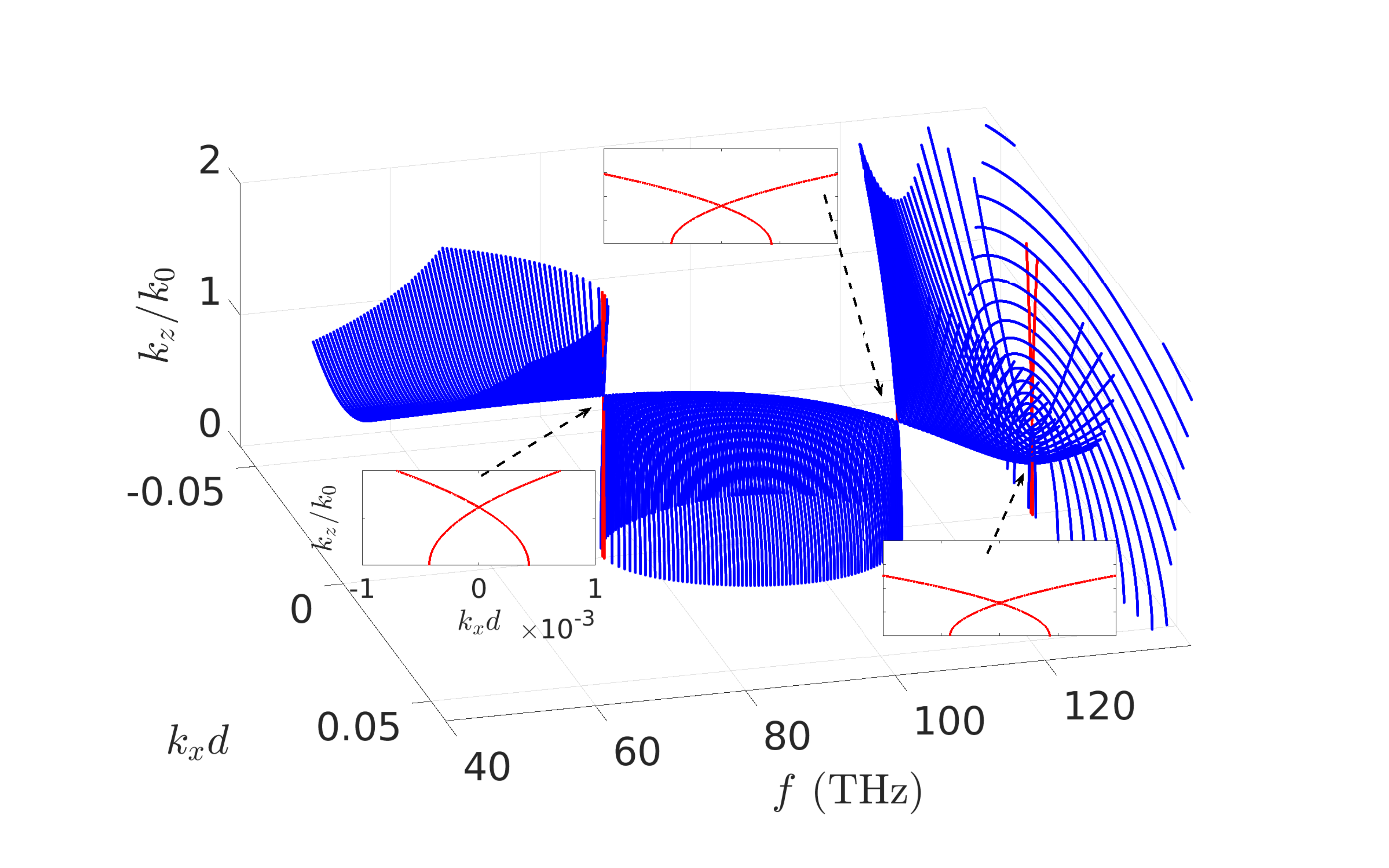}
  \includegraphics[scale=0.35]{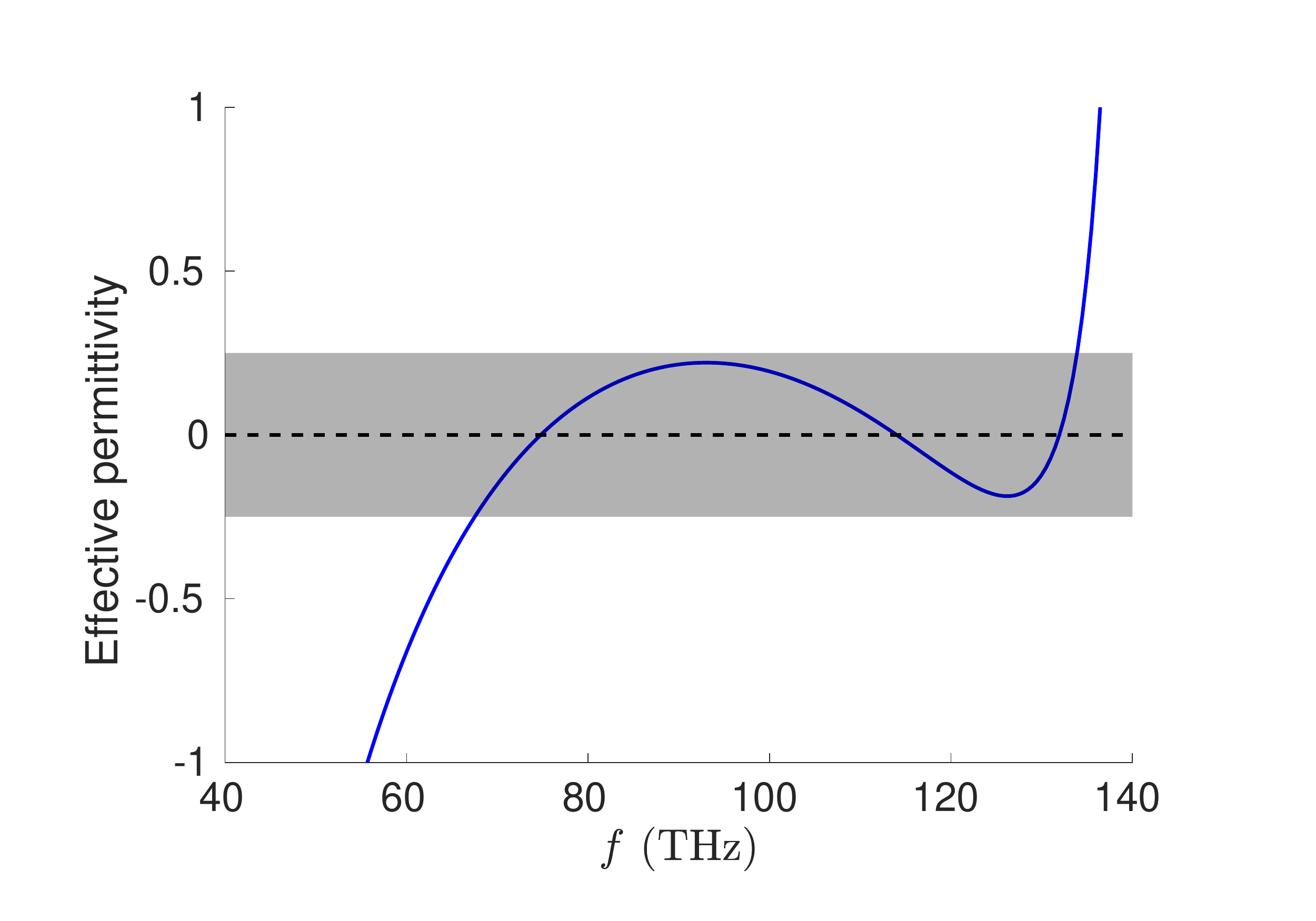}
  \caption{A two-oscillator Lorentz material is used as a host for the
    graphene plasmonic crystal. Upper: Dispersion relation $k_x(k_z)$ as a
    function of frequency; the Dirac cones are marked in red and also shown
    in inset. Lower: Effective permittivitity; the ENZ area is shaded in
    grey.}
  \label{fig:bands__multiOsc}
\end{figure}

We can further generalize our approach by assuming dielectric materials
with permittivity given by   a multi-oscillators Lorentz model \cite{solidStateBook} where many
interband transitions take place and drastically contribute to the
dielectric function. Such a material provides many Dirac points in the
${\boldsymbol k}$ space. Subsequently, the $\varepsilon_z^\text{eff}$
becomes zero many times and thus, the ENZ frequency range is significantly
expanded. We propose this hypothetical scenario due to the great progress
of the material engineering where  new materials with desirable electronic
properties can be designed in purpose.  The Lorentz model that includes
multiple oscillators is
\begin{equation}
  \label{eq:LorentzMultip}
  \varepsilon(\omega) = \varepsilon_\infty + \sum_i^N \frac{g_i\
  \omega_{0i}^2}{\omega_{0i}^2 - \omega^2 +i\omega\Gamma_i},
\end{equation}
where $N$ is the number of the oscillators each one has resonance frequency
$\omega_{0i}$, damping factor $\Gamma_i$, and strength $g_i$. Here, we
consider a two-oscillator ($N=2$)  Lorentz material  with the parameters:
$\varepsilon_\infty=4$, $(g_1, g_2) = (0.34,~ -2.4)$, and
$(\omega_{01},\omega_{02})= (0.59, 0.71 )$ eV. Thence, we calculate the
dispersion relation in (${\boldsymbol k}~,\omega)$ and present the results
in Figure \ref{fig:bands__multiOsc} (upper) where we observe three Dirac
points at the frequencies $(75,~114,~132)$ THz. In the lower panel of the
same figure we outline the effective permittivity in frequency where the
ENZ band has been significantly expanded in a range over than $60$ THz that
is approximately eight times wider range compared to the example of a
constant permittivity host material.

\section{Conclusion}

Metamaterials are artificial structures that manipulate electromagnetic
waves on a sub-wavelength scale and provide optical properties that are not
found in nature under normal conditions, such as negative refraction and
epsilon-near-zero behavior. ENZ metamaterials exhibit exotic properties
that include wave propagation through very narrow channels without
diffraction and phase delay. Plasmonic crystals are a particular class of
metamaterials that consist of periodically arranged metallic layers with
distances shorter than the (free space) wavelength. Graphene-based
plasmonic crystals are tunable metamaterials since their optical properties
can be dynamically tuned by controlling the operating frequency and the
electronic doping in graphene. Fine tuning a graphene-based plasmonic
crystal yields ENZ behavior in terahertz and infrared frequencies. The
appearance of ENZ condition coincides with a plasmonic Dirac cone in the
wavenumber space and thus, Dirac points determine the neighborhood where
the metamaterial behaves as an ENZ media.

Using an ordinary dielectric as the host material of the plasmonic crystal
yields a single Dirac point restricting the ENZ effect to a very narrow
frequency range. This may be a critical issue for for real applications
that typically involve a range of operating frequencies forming a
wavepacket. In this paper we proposed a modified design that expands the
frequency range of the ENZ behavior. We replaced the host dielectric in the
plasmonic crystal with a Lorentz dispersive material which has a frequency
dependent permittivity. We demonstrated that the combination of doped
graphene and Lorentz materials yield multiple Dirac points and,
subsequently, achieved an ENZ effect over a relatively large frequency
range.

\begin{acks}
  We acknowledge partial support from EFRI 2-DARE NSF Grant No. 1542807
  (MMat, EK); and ARO MURI Award No. W911NF14-0247 (MMat, MMai, EK). This
  project was partially supported by the University of Minnesota's
  Undergraduate Research Opportunities Program (WXB).
\end{acks}



\end{document}